\documentclass[aps,prl,reprint,superscriptaddress]{revtex4-1}
\usepackage{graphicx}
\usepackage{rotating}
\usepackage{amssymb}    
\usepackage{amsmath}   
\usepackage{epsfig}
\usepackage[normalem]{ulem}   
\usepackage{bm}   
\usepackage{color}

\date{\today}

\begin{document}

\title{Parameter Uncertainties on the Predictability of Periodicity and Chaos}

\author{E. S. Medeiros}
\email{esm@if.usp.br}
\affiliation{Institute of Physics, University of S\~ao Paulo, Rua do Mat\~ao, Travessa R 187, 05508-090, S\~ao Paulo, Brazil}
\affiliation{Institute for Complex Systems and Mathematical Biology, SUPA, University of Aberdeen, AB24 3UE Aberdeen, United Kingdom}
\author{I. L. Caldas}
\affiliation{Institute of Physics, University of S\~ao Paulo, Rua do Mat\~ao, Travessa R 187, 05508-090, S\~ao Paulo, Brazil}
\author{M. S. Baptista}
\affiliation{Institute for Complex Systems and Mathematical Biology, SUPA, University of Aberdeen, AB24 3UE Aberdeen, United Kingdom}

\begin{abstract}

Nonlinear dynamical systems, ranging from insect populations to lasers and chemical reactions, might exhibit sensitivity to small perturbations in their control parameters, resulting in uncertainties on the predictability of tunning parameters that lead these systems to either a chaotic or a periodic behavior. By quantifying such uncertainties in four different classes of nonlinear systems, we show that this sensitivity is to be expected because the boundary between the sets of parameters leading to chaos and to periodicity is fractal. Moreover, the dimension of this fractal boundary was shown to be roughly the same for these classes of systems. Using an heuristic model for the way periodic windows appear in parameter spaces, we provide an explanation for the universal character of this fractal boundary.

\end{abstract}

\maketitle

The topology of solutions of nonlinear dynamical systems can be severely affected by small perturbations in their control parameters \cite{Kuznetsov2004}. The so-called parameter sensitivity has been experimentally observed in dynamical and nonlinear models of systems in different areas of knowledge \cite{Costantino1997, Ren1997, Lindberg2007, Kolokolov2013}. The cause of this sensitivity is the existence of bifurcations, such as the ones leading to crisis \cite{Grebogi1983} where chaotic attractors abruptly bifurcate into periodic ones (or vice-versa). The most profund consequence of this parameter sensitivity is to limit the ability of someone to set a parameter of a system that surely places it into either a chaotic or a periodic behavior.

In parameter spaces and bifurcation diagrams of discrete and continuous-time nonlinear dynamical systems, the set of parameters leading to chaotic behavior is intertwined with hierarchical structures of sets of parameters leading to periodic stable behavior, the complex periodic windows (CPWs) \cite{Murilo2003,Façanha2013}. The structure of CPWs describes a scenario for the way that periodicity and chaos appear in a large variety of nonlinear dynamical systems: lasers \cite{Bonatto2005}, electronic circuits \cite{Albuquerque2012}, population dynamics \cite{Slipantschuk2010}, nonlinear oscillators \cite{Medeiros2013}, etc \cite{Medeiros2010, Murilo1996,Gallas2010, Bonatto2008b}. These periodic structures were numerically shown to have self-similar-like properties, i.e., the structure of the CPWs is preserved for any scale of the parameter space \cite{Glass1983}. Moreover, the CPWs appear aligned in infinite torsion and period-adding sequences \cite{Bonatto2007, Medeiros2013,Bonatto2008b}. Recently, many 
researchers have been carrying out additional theoretical, experimental, and numerical works to find mechanisms to explain the existence, the genesis, and the organization of CPWs. Important results were found in two-dimensional parameter spaces of dynamical systems for which the Shilnikov theorem \cite{Gaspard1982, Gaspard1984} can be applied, and homoclinic orbits converge to saddle-focus equilibrium points. It was found that CPWs are connected to each other forming spiral-like structures emerging from the homoclinic bifurcation points, i.e, the parameters corresponding to the saddle-focus for which homoclinic orbits converge. Moreover, sets of homoclinic bifurcation points are aligned forming homoclinic bifurcation curves. From each point in these curves an entire spiral-like structures emerge \cite{Rene2010,Vitolo2011,Barrio2011,Barrio2012,Bonatto2008,Holokx2008}. This configuration is suggesting that these spiral-like structures are accumulating in a fractal way in two-dimensional parameter spaces. 
These spiral-like structures of CPWs have also been observed in real-world experiments \cite{Maranhao2008,Stoop2010,Cabeza2013}. Additionally, in Ref. \cite{Hunt1997} it has been argued that the width (Lebesgue measure) of the CPWs decreases exponentially with the period of the attractor and the topological entropy of the surrounding chaotic region [See supplementary material].

Parameter sets corresponding to CPWs not only appear in all scales of parameter spaces (self-similar), but they also have positive {\it Lebesgue measure}. Self-similar sets with non-zero Lebesgue measure are called {\it fat Cantor} sets. These sets are topologically equivalent to the usual {\it Cantor set}, but their properties are different, specially, their capacity dimension. In the case of fat Cantor sets, the capacity dimension is equal to the dimension of the embedding euclidean space \cite{Umberger1986,Farmer1985}. Therefore, the dimension of CPWs in two-dimensional parameter spaces would be $D=2$. CPWs form fat Cantor sets. Consequently, the likelihood of CPWs being experimentally found in any scale of parameter spaces is high. In fact, both periodicity and chaos in the asymptotic limit are likely to be found by either making a controlled tunning of the parameters or by taking a random sample of parameters. However, if CPWs are self-similar, it could lead to uncertainties for the predictability of 
tunning the parameters to produce either periodic or chaotic behaviors, since that, even though possessing integer dimensions, self-similar sets can have fractal boundaries. Consequently, if CPWs of a nonlinear dynamical system have fractal boundaries, predictability in the setting of parameters that would surely take the system to either a periodic or a chaotic behavior could be severely compromised.

Even though chaotic regions appearing in many bifurcation diagrams of one-dimensional systems \cite{Jakobson1981} are known to be self-similar fat cantor sets, the kind of self-similarity present in CPWs appearing in two-dimensional parameter spaces was so far an open problem. Visual inspections of sucessive enlargements of parameter spaces regions \cite{Gallas1993,Lorenz2008} have suggested that CPWs are self-similar. Recently new evidences are pointing out that CPWs are self-similarly organized, occurring for parameters in the center of spiral-like structures. \cite{Gaspard1982, Gaspard1984,Rene2010,Vitolo2011,Barrio2011,Barrio2012}.

In this work, we indeed show that CPWs are self-similar. Self-similarity appears not only in the parameter widths of the CPWs but also in the boundaries between them and the chaotic regions. We show that the functionality of the self-similarity of CPWs regarding their widths can be both, power-law or even exponential as proposed in \cite{Hunt1997} and observed in \cite{Viana2010}. We numerically estimate the capacity dimension of the boundaries between parameters corresponding to CPWs and parameters leading to chaos, showing that they are usual {\it skinnies fractals} possessing a non-integer exterior capacity dimension. Consequently, in any scale in two-dimensional parameter spaces of a large class of nonlinear systems, there are always uncertainties associated with the predictability for tunning the parameters that surely lead the system to either a periodic or a chaotic behavior. In our simulations, the  capacity dimension of these boundaries seem to be universal for different classes of nonlinear 
dynamical systems. We then developed an heuristic model for the appearance of CPWs and showed that the self-similarity of the widths of the CPWs appearing in this model produce fractal boundaries. This was quantified by the relation between the capacity dimension of the parameter boundaries and the decreasing rate of CPWs widths along accumulating sequences. The capacity dimension of the boundaries predicted by the model agrees with the values obtained in our simulations. So, the universal character of these boundaries is attributed to the way CPWs appear. They appear in sequences organized by their "order" and have parameter widths that decay as a power-law with their order \cite{Nota1}.

To calculate the dimension of the boundary of two sets, a special capacity dimension has been defined. Defining the set $S$ as a boundary between two regions, considering {\bf $S$}($\varepsilon$) a new set formed by all points within a distance $\varepsilon$ from {\bf $S$}, then, defining $\bar S(\varepsilon)= S(\varepsilon) - S$, the {\it exterior capacity dimension} $d_x$ \cite{Grebogi1985} was defined by: 
\begin{eqnarray}
 d_x = \displaystyle \lim_{\varepsilon \to 0}{\frac{\ln V[\bar S(\varepsilon)]}{\ln \varepsilon}},
\end{eqnarray}
where $V[\bar S(\varepsilon)]$ is the volume of the set $\bar S(\varepsilon)$. This operation is however difficult to be calculated directly. An alternative way is done by considering the {\it uncertain exponent}.

Along a direction transversal to $S$, we take three parameter values $\varepsilon$-distant to each other. Count the number of uncertain triplets, i.e., we take three neighboring parameter values and check whether these parameters do not lead to an unique type of behavior (chaos or periodicity). From the uncertain parameters one can calculate the uncertain fraction $f(\varepsilon)$ of parameters \cite{Grebogi1985}. It has been verified for certain one-dimensional quadratic maps that the uncertain fraction $f(\varepsilon)$ varies as a power-law with $\varepsilon$, i.e., $f(\varepsilon)=K\varepsilon^{\alpha}$, where the factor $K$ was believed to be dependent of the parameter range considered, and the exponent $\alpha$ has been called uncertainty exponent, and believed to be independent of the considered parameter interval \cite{Grebogi1985}. 

The uncertainty exponent can be directly related to the exterior dimension of a set \cite{McDonald1984}. It has been heuristically demonstrated by considering $N(\gamma)$ as the minimum number of $D$-dimensional cubes of side $\gamma$ required to cover the boundary of the set \cite{McDonald1984}. It is well-known that $N(\gamma)$ scales with the cube side $\gamma$ as $N(\gamma)=\gamma^{-d_x}$. Setting $\gamma$ to be equal to the parameter error $\varepsilon$, the uncertain region will be of the order of the total volume of all $N(\varepsilon)$ $D$-dimensional cubes required to cover the boundary. The volume of one uncertain cube is given by $\varepsilon^D$, so the total uncertain volume is given by $\varepsilon^DN(\varepsilon) \sim \varepsilon^{D-d_x}$. Assuming that the uncertain fraction is proportional to the uncertain volume, then:
\begin{eqnarray}
 d_x=D-\alpha.
 \label{eq:dim}
\end{eqnarray}

Our numerical results are based on simulations of four different classes of dynamical systems. We consider parameter spaces regions for which the parameters corresponding to CPWs are hierarchically distributed giving self-similar features to the region.

We consider the R\"ossler oscillator for which the Shilnikov theorem can be applied. This system is described by the following set of nonlinear differential equations:
\begin{equation}
     \begin{split}
    \dot{x}& = -y-z,\\
    \dot{y}& = x+ay,\\
    \dot{z}& = (b+z)x-cz.
  \end{split}
  \label{chemicaloscillator}
\end{equation}
We investigate an extension of the parameter plane $a \times c$, where the complex periodic structures emerge from homoclinic orbits and are spiral-shaped organized in sequences. The other parameter $b$ of Eq.~(\ref{chemicaloscillator}) is fixed at $b=0.3$ \cite{Gallas2010}.

The class of nonlinear forced oscillators are represented by the Morse oscillator which is governed by the following nonlinear differential equation \cite{Scheffczyk1991}:   
\begin{equation} 
\ddot{x}+d\dot{x}+8e^{-x}(1-e^{-x}) = 2.5\cos(\omega t),
\label{morseoscillator}
\end{equation}
in the parameter plane $\omega \times d$, the CPWs are aligned in sequences of period and torsion-adding.

We also work with a loss-modulated CO$_2$ laser described by a rate-equation with a time-dependent parameter:
\begin{equation}
    \begin{split}
    \dot{u}& = \frac{1}{\tau}(z-k(t))u,\\
    \dot{z}& = (z_0-z)\gamma-uz,
     \end{split}
  \label{co2laser}
\end{equation}
where $k(t)=k_0(1+a\cos2\pi ft)$. We investigate a complex periodic sequence in the $a \times f$ parameter plane. All other parameters are fixed: $\tau=3.5\times 10^{⁻9}$ $s$, $\gamma=1.978 \times 10^5$ $s^{-1}$, $z_0=0.175$, and $k_0=0.1731$ \cite{Bonatto2005}. 

Finally, we consider a sequence of CPWs in the $K \times \omega$ parameter spaces of the well-known circle map described by the discrete-time equations \cite{Perbak1983}:
\begin{equation}
    x_{n+1} = x_n + \omega - \frac{K}{2\pi}\sin(2\pi x_n),
    \label{henonmap}
\end{equation}

In the two-dimensional parameter spaces of those systems, we select $3.0 \times 10^4$ pairs ($a_0$, $b_0$) of random parameters uniformly space distributed and compute the largest Lyapunov exponents to determine if the correspondent state is periodic ($\lambda<0$) or chaotic ($\lambda>0$). Then, each pair of parameters is perturbed by an error $\varepsilon$ in both orientations along one parameter. This process generates $6.0 \times 10^4$ pairs ($a_0 \pm \varepsilon$, $b_0$) of parameters. We also obtain the Lyapunov exponent of states corresponding to the perturbed parameter pairs. We compare only the unperturbed chaotic parameters (that produces chaotic attractors, i.e., $\lambda>0$) to their two correspondent perturbed pairs along the horizontal direction. If at least one of them is not chaotic, the pair ($a_0$, $b_0$) is counted as an uncertain pair for the error value $\varepsilon$. We record the uncertain fraction for an error interval. 

\begin{figure}[!htp] \centering
\includegraphics[width=8.5cm,height=10cm]{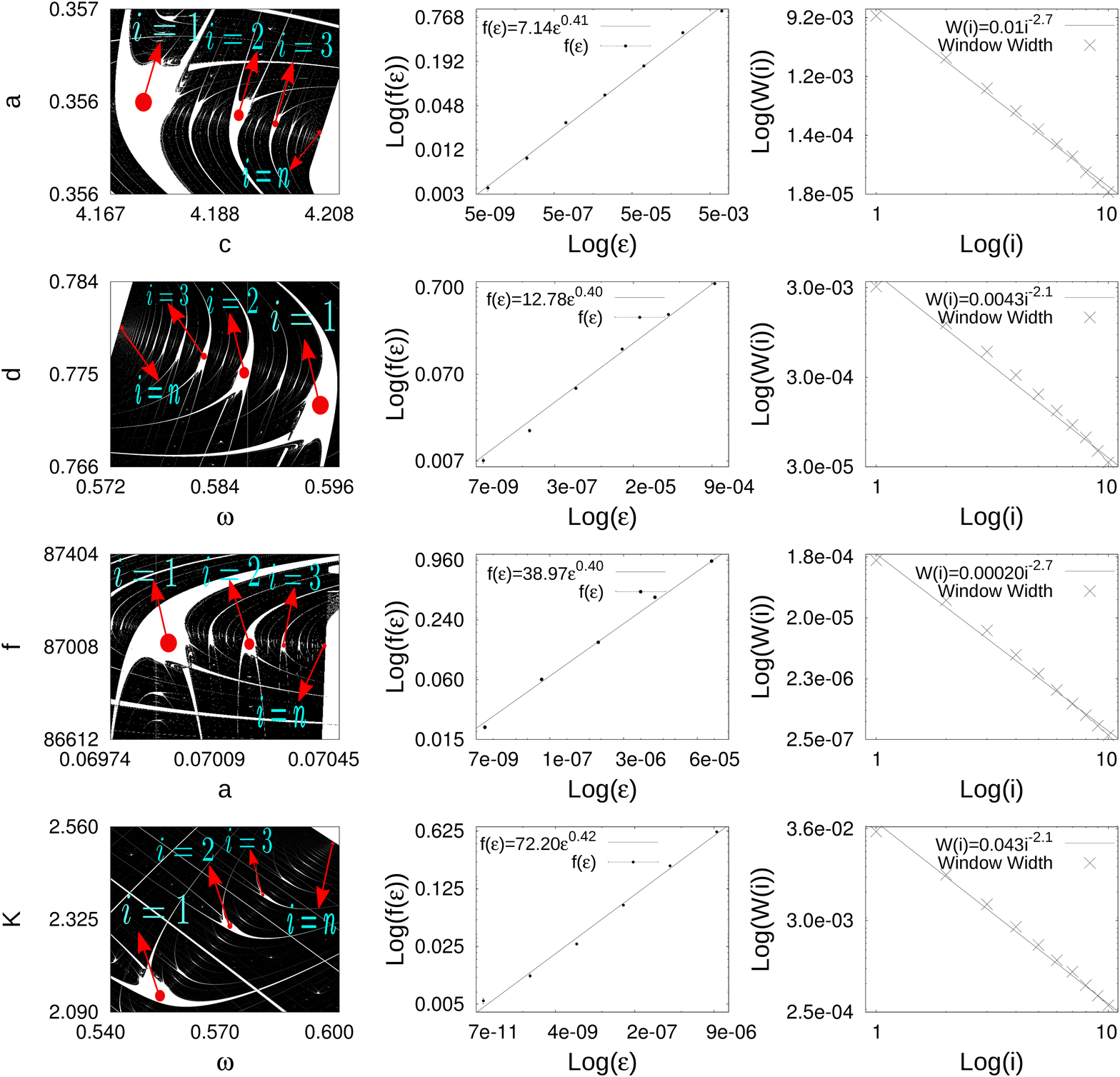}
\caption{(Left) Two-dimensional parameter spaces of the four considered dynamical systems. Black regions represent the chaotic parameter set. White regions represent the periodic parameter set. (Center) The uncertain fraction $f(\varepsilon)$ of the chaotic sets shown in (Left) scales as power law with the error $\varepsilon$. (Right) The width of the periodic windows shown in (Left) versus their order $i$. The window width is obtained by taking an one-dimensional cut of the two-dimensional parameters space and measuring the distance from the initial saddle-node bifurcation to the final crisis of a peridic windows \cite{Yorke1985}.}
\label{figure1}
\end{figure}

In Figure \ref{figure1}(Left), we show the parameter spaces for the four systems considered, for which $f(\varepsilon)$ and consequently $d_x$ are calculated. The black regions indicate the set of parameters leading to chaos, while the white regions correspond to parameters leading to periodic stable behavior. In these figures, CPWs are aligned along sequences accumulating in periodic regions of parameter spaces. Here, a specific sequence of CPWs is ordered by its characteristic period-adding rule. The CPWs where attractors have theirs periods added along the sequence are identified by their order $1,2,3,...$. The window with the largest width of a sequence has the lowest order, $i=1$. For dynamical systems where torsion and rotation numbers are defined, sequences can also be identified by those parameters where frequency locking occurs \cite{Medeiros2013}. There exists infinite sequences with different period-adding rules and CPWs sizes. For this work, we measure the width of periodic windows beloging only 
to the main sequences (larger width) identified by the filled circles in Fig. \ref{figure1}(Left). In Figure \ref{figure1}(Center), for the correspondent parameter space shown in Fig. \ref{figure1}(Left), we show the fraction $f(\varepsilon)$ of uncertain periodic parameters as a function of the error $\varepsilon$. The straight line is a power-law fitting between $f(\varepsilon)$ and $\varepsilon$ which provides the uncertainty exponent $\alpha$. We observe that the exponent $\alpha$ is in the same confidence interval given by $\alpha = 0.40 \pm 0.04$ for the different classes of dynamical systems considered here. The standard deviation of $\alpha$ has been obtained by considering that the occurrence of uncertain parameters are random events. From $\alpha$ in Eq. (\ref{eq:dim}) we obtain that the dimension of the boundary between the chaotic and periodic parameter sets is given by $d_x=1.60 \pm 0.04$. In Figure \ref{figure1}(Right), we show the width of the first $10$ CPWs as a function of the order it 
appears along the main accumulation sequence. We fit a power-law of the form $W(i)=A/i^l$ to the way that CPW widths decrease \cite{Nota1}. 

The measurements of Fig. \ref{figure1} indicate that the exterior capacity dimension of the boundaries between parameters corresponding to CPWs and parameters leading to chaos is universal for different classes of dynamical systems. To understand why that would be so, we formulate an heuristic model for the appearance of CPWs and chaotic regions where the observed decreasing of the CPWs width is considered. In our model, CPWs are created by removing pieces of a one-dimensional cut of the two-dimensional parameters space, Fig. $2$. The gaps represent a CPW and the remaining intervals represent chaotic regions. One begins with a chaotic interval of length $L_0$ and removes the amount correspondent to the width $W_1$ of a CPW of first order, $W_1=AL_0/1^l$, leaving a chaotic interval of length $L_1=L_0-AL_0/1^l$. Next, one removes from $L_1$ the amount correspondent to the width of a CPW of second order, $W_2=AL_0/2^l$, leaving a chaotic interval of length $L_2=L_1-AL_0/2^l$. Continuing in this manner, {\it ad 
infinitum}, we will obtain a set of elements corresponding to the chaotic parameters from which the intervals corresponding to CPWs of one sequence have been removed. To decide if this asymptotic set is a fat fractal, we obtain its uncertainty exponent $\alpha_M$, the set will be a fat fractal for $0<\alpha_M \leqslant 1$ \cite{Umberger1986,Farmer1985}. Moreover, for one-dimensional quadratic maps, the parameters corresponding to chaotic behavior have been demonstrated to have nonzero Lebesgue measure (fat fractal) \cite{Jakobson1981}.

\begin{figure}[!htp] \centering
\includegraphics[width=6.5cm,height=3.5cm]{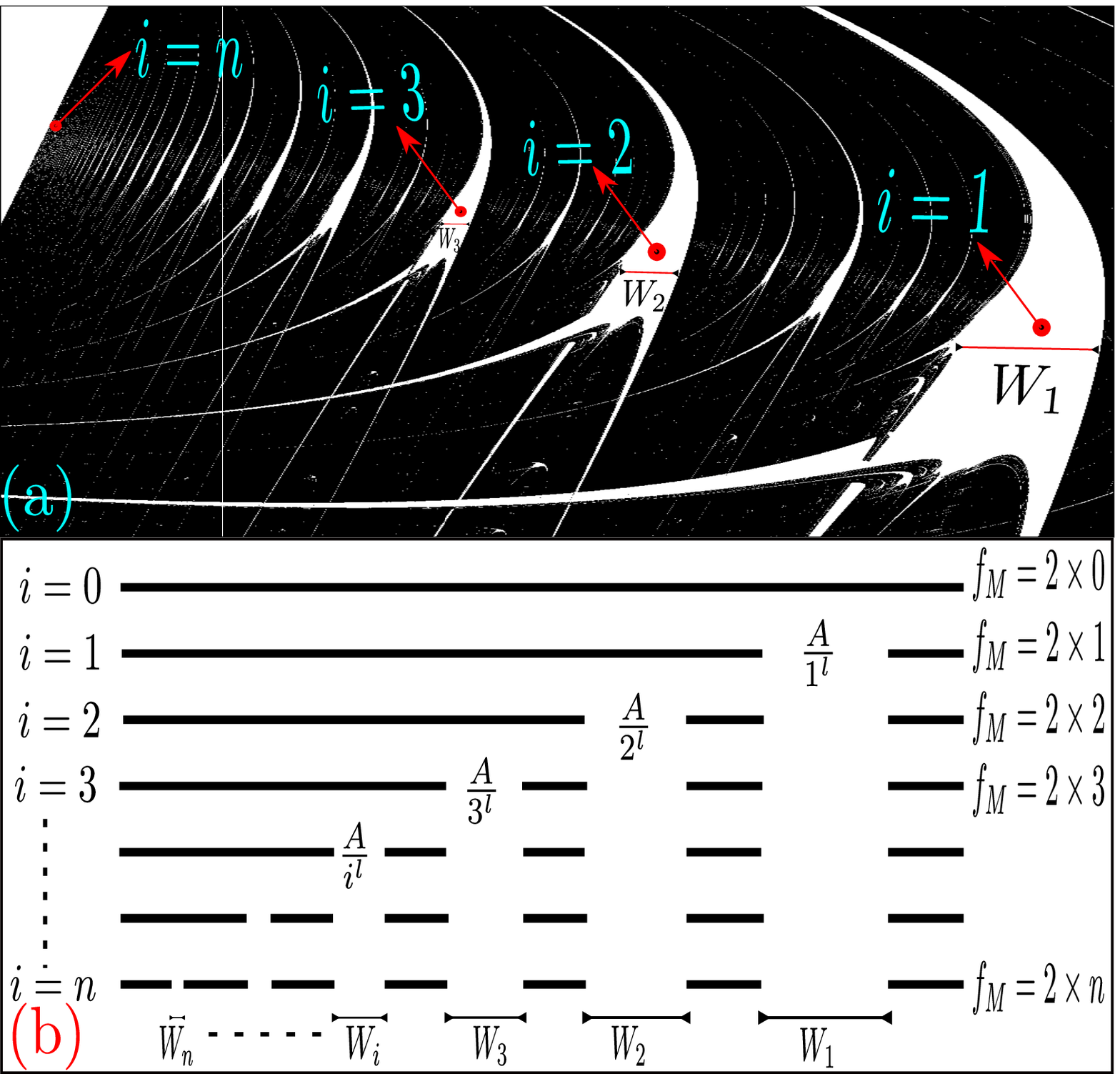}
\caption{(a) A two-dimensional parameter space, the white regions represent parameters corresponding to CPWs. (b) Schematic of the formation of CPWs that have fractal boundaries. For $i=1$ and $l=2.2$, a periodic window is created by the removal of $A/1^{2.2} = A$ of the initial line. For $i=2$, a periodic window is created by removal $A/2^{2.2} \simeq A/0.22$ of the line, and so on. The remaining pieces of the line represent chaotic regions. Secondary accumulation of CPWs can be created by removing two intervals at each iteration. $f_M$ is the number of uncertain points of each iteration.}
\label{figure2}
\end{figure}

In order to obtain the uncertainty exponent of the heuristic model, we count the number of uncertain parameters $f_M(i)$ as the CPWs width $W$ decreases. The number of uncertain parameters of each iteration is given by $f_M(i)=2i$, a number representing the numbers of boundaries between the remaining pieces and the gaps. The width of the CPW decreases at each iteration by $W_i = A/i^{l}$, Fig. \ref{figure2}. The uncertain fraction of parameter $f(\varepsilon)$ is written as function of the error $\varepsilon$, i.e., $f(\varepsilon) \propto \varepsilon^{\alpha}$. So, the number of uncertain parameters of the model can be written as $f_M(i) \propto \varepsilon_{M}^{\alpha_M}$. Considering that the model error $\varepsilon_M$ is inversely proportional to the CPW width, $\varepsilon_M \propto i^l$, using that $f_M(i)=2i$, we derive the uncertainty exponent, $\alpha_M$, of the heuristic model:
\begin{equation} 
\alpha_M = \displaystyle \lim_{i \to \infty}\frac{\log(2i)}{\log(i^l)}=\frac{1}{l}.
\label{uncertexpmodel}
\end{equation}
In this heuristic model we consider only the main sequence of peridic windows, numbered in Fig. (\ref{figure1}). We believe that the uncertain exponent, $\alpha_M$, is independent of the number of sequences considered, once that, for a larger number of sequences the fraction of uncertain elements, $f_M(\varepsilon)$, per iteration is higher, but the error, $\varepsilon$, will be lower, so the limit of Eq. (\ref{uncertexpmodel}) must be the same despite of the number of sequences considered by the model. In fact, we obtain that $\alpha_M = \alpha$ (see Table \ref{table1}), an evidence that smaller sequences do not contribute much for $\alpha_M$. Using Eq. (\ref{eq:dim}) in Eq. (\ref{uncertexpmodel}), we obtain the exterior capacity dimension $d_{Mx}$ of the model:
\begin{equation} 
d_{Mx}=\frac{lD-1}{l},
\label{analiticaldim}
\end{equation}
where $l$ is the exponent of the width decreasing and $D$ is the dimension of the embedding euclidean space (for two-dimensional parameters space, $D=2$). Substituting the values of $l$, obtained in Fig. \ref{figure1}, in Eq. (\ref{analiticaldim}), we obtain the exterior capacity dimension expected for the boundary between CPWs and chaos according to the heuristic  model. The standard deviations of $l$, $\alpha_M$ and $d_{Mx}$ are obtained by propagating the uncertainties of the window width measured in bifurcations diagrams. 

In Table \ref{table1}, for all dynamical systems considered here, we compare the measurements from the simulations with the results provided by the model when the measured exponents $l$ for the decreasing of CPWs are given. We verify that the exterior capacity dimension $d_{Mx}$ provided by the model agrees with the exterior capacity dimension obtained in our simulations shown in Fig. \ref{figure1}.

\begin{table*}[!htp]
\centering
\begin{tabular}{|p{4cm}|p{2.0cm}|p{2.3cm}|p{2.3cm}|p{2.3cm}|p{2.3cm}|}
\hline
{\bf Dynamical System} & {\bf $l$} & {\bf $d_x=D-\alpha$} & {\bf $d_{Mx=(lD-1)/l}$} & {\bf $\alpha$} & {\bf $\alpha_M$}\\
\hline\hline
R\"ossler Oscillator & $2.7 \pm 0.1$ & $1.59 \pm 0.04$ & $1.63 \pm 0.06$ & $0.41 \pm 0.04$ & $0.37 \pm 0.01$\\
\hline
Morse Oscillator & $2.1 \pm 0.1$ & $1.60 \pm 0.04$ & $1.52 \pm 0.07$ & $0.40 \pm 0.04$ & $0.48 \pm 0.02$\\
\hline
CO$_2$ Laser & $2.7 \pm 0.1$ & $1.60 \pm 0.04$ & $1.63 \pm 0.06$ & $0.40 \pm 0.04$ & $0.37 \pm 0.01$\\
\hline
Circle Map & $2.1 \pm 0.1$ & $1.58 \pm 0.04$ & $1.52 \pm 0.07$ & $0.42 \pm 0.04$ & $0.48 \pm 0.02$ \\
\hline
\end{tabular}
\caption{In this table, we show in the first column the value of the exponent $l$ obtained from the fitting in Fig. \ref{figure1}(right), in the second column the values of $d_x$ calculated from the data in Fig. \ref{figure1}(center), and in the third column, the values of $d_Mx$ using Eq.~(\ref{analiticaldim}). In fourth column, the values of $\alpha$ and fifth column the values of $\alpha_M$.}
\label{table1}
\end{table*}

In conclusion, we have shown that the boundaries between parameters corresponding to CPWs and parameters leading to chaos in two-dimensional parameter spaces are fractals. Consequently, the ability of tunning a parameter of a nonlinear dynamical system to set its behavior to be surely either chaotic or periodic is seriously compromised by this fine structure of the boundary in all scales of the parameter space. In our simulations, we found that the capacity dimension of such boundaries seems to be universal for different classes of dynamical systems treated in this work. From an heuristic model for the appearance of CPWs, we deduce a relation between the exterior capacity dimension and the exponent of decreasing of the CPWs widths along sequences. The dimension deduced from the model agrees with values observed in our simulations. This fact strongly suggests that the universality observed for this boundary between chaos and periodicity is a consequence of the power-law fashion with which periodic windows 
decrease their sizes as a function of their order. We also remark that the decreasing of the width of periodic structures immersed in parameters corresponding to quasi-periodic behavior, called Arnold tongues \cite{Glass1982}, has been observed to have a power-law dependence on its period \cite{Ecke1989}. These facts give support to our main claim that the decreasing of the window width of the CPWs in period-adding sequences can be described by a power-law function. However, CPWs can also have their sizes exponentially decreasing with their order, when the accumulation of CPWs forms the spiral-like structures due to the homoclinic bifurcation scenario (see supplementary material). These spiral-like structures are however not predominant all over the domain of the parameter space, but coexist with accumulation sequences where CPWs have their widths decreasing in a power-law fashion. 

We would like to thank the partial support of this work by the Brazilian agencies FAPESP (process: 2011/19296-1), CNPq and CAPES.


\begin{thebibliography}{45}%
\makeatletter
\providecommand \@ifxundefined [1]{%
 \@ifx{#1\undefined}
}%
\providecommand \@ifnum [1]{%
 \ifnum #1\expandafter \@firstoftwo
 \else \expandafter \@secondoftwo
 \fi
}%
\providecommand \@ifx [1]{%
 \ifx #1\expandafter \@firstoftwo
 \else \expandafter \@secondoftwo
 \fi
}%
\providecommand \natexlab [1]{#1}%
\providecommand \enquote  [1]{``#1''}%
\providecommand \bibnamefont  [1]{#1}%
\providecommand \bibfnamefont [1]{#1}%
\providecommand \citenamefont [1]{#1}%
\providecommand \href@noop [0]{\@secondoftwo}%
\providecommand \href [0]{\begingroup \@sanitize@url \@href}%
\providecommand \@href[1]{\@@startlink{#1}\@@href}%
\providecommand \@@href[1]{\endgroup#1\@@endlink}%
\providecommand \@sanitize@url [0]{\catcode `\\12\catcode `\$12\catcode
  `\&12\catcode `\#12\catcode `\^12\catcode `\_12\catcode `\%12\relax}%
\providecommand \@@startlink[1]{}%
\providecommand \@@endlink[0]{}%
\providecommand \url  [0]{\begingroup\@sanitize@url \@url }%
\providecommand \@url [1]{\endgroup\@href {#1}{\urlprefix }}%
\providecommand \urlprefix  [0]{URL }%
\providecommand \Eprint [0]{\href }%
\providecommand \doibase [0]{http://dx.doi.org/}%
\providecommand \selectlanguage [0]{\@gobble}%
\providecommand \bibinfo  [0]{\@secondoftwo}%
\providecommand \bibfield  [0]{\@secondoftwo}%
\providecommand \translation [1]{[#1]}%
\providecommand \BibitemOpen [0]{}%
\providecommand \bibitemStop [0]{}%
\providecommand \bibitemNoStop [0]{.\EOS\space}%
\providecommand \EOS [0]{\spacefactor3000\relax}%
\providecommand \BibitemShut  [1]{\csname bibitem#1\endcsname}%
\let\auto@bib@innerbib\@empty
%</preamble>
\bibitem [{\citenamefont {Kuznetsov}(2004)}]{Kuznetsov2004}%
  \BibitemOpen
  \bibfield  {author} {\bibinfo {author} {\bibfnamefont {Y.~A.}\ \bibnamefont
  {Kuznetsov}},\ }\href@noop {} {\emph {\bibinfo {title} {Elements of applied
  bifurcation theory}}}\ (\bibinfo  {publisher} {Springer},\ \bibinfo {address}
  {United States of America},\ \bibinfo {year} {2004})\BibitemShut {NoStop}%
\bibitem [{\citenamefont {Costantino}\ \emph {et~al.}(1997)\citenamefont
  {Costantino}, \citenamefont {Desharnais}, \citenamefont {Cushing},\ and\
  \citenamefont {Dennis}}]{Costantino1997}%
  \BibitemOpen
  \bibfield  {author} {\bibinfo {author} {\bibfnamefont {R.~F.}\ \bibnamefont
  {Costantino}}, \bibinfo {author} {\bibfnamefont {R.~A.}\ \bibnamefont
  {Desharnais}}, \bibinfo {author} {\bibfnamefont {J.~M.}\ \bibnamefont
  {Cushing}}, \ and\ \bibinfo {author} {\bibfnamefont {B.}~\bibnamefont
  {Dennis}},\ }\href@noop {} {\bibfield  {journal} {\bibinfo  {journal}
  {Science}\ }\textbf {\bibinfo {volume} {275}},\ \bibinfo {pages} {389}
  (\bibinfo {year} {1997})}\BibitemShut {NoStop}%
\bibitem [{\citenamefont {Ren}\ \emph {et~al.}(1997)\citenamefont {Ren},
  \citenamefont {Hu}, \citenamefont {Zhang}, \citenamefont {Wang},
  \citenamefont {Gong},\ and\ \citenamefont {Xu}}]{Ren1997}%
  \BibitemOpen
  \bibfield  {author} {\bibinfo {author} {\bibfnamefont {W.}~\bibnamefont
  {Ren}}, \bibinfo {author} {\bibfnamefont {S.~J.}\ \bibnamefont {Hu}},
  \bibinfo {author} {\bibfnamefont {B.~J.}\ \bibnamefont {Zhang}}, \bibinfo
  {author} {\bibfnamefont {F.~Z.}\ \bibnamefont {Wang}}, \bibinfo {author}
  {\bibfnamefont {Y.~F.}\ \bibnamefont {Gong}}, \ and\ \bibinfo {author}
  {\bibfnamefont {J.~X.}\ \bibnamefont {Xu}},\ }\href@noop {} {\bibfield
  {journal} {\bibinfo  {journal} {Int. J. of Bif. and Chaos}\ }\textbf
  {\bibinfo {volume} {7}},\ \bibinfo {pages} {1867} (\bibinfo {year}
  {1997})}\BibitemShut {NoStop}%
\bibitem [{\citenamefont {Valling}\ \emph {et~al.}(2007)\citenamefont
  {Valling}, \citenamefont {Krauskopf}, \citenamefont {Fordell},\ and\
  \citenamefont {Lindberg}}]{Lindberg2007}%
  \BibitemOpen
  \bibfield  {author} {\bibinfo {author} {\bibfnamefont {S.}~\bibnamefont
  {Valling}}, \bibinfo {author} {\bibfnamefont {B.}~\bibnamefont {Krauskopf}},
  \bibinfo {author} {\bibfnamefont {T.}~\bibnamefont {Fordell}}, \ and\
  \bibinfo {author} {\bibfnamefont {A.~M.}\ \bibnamefont {Lindberg}},\
  }\href@noop {} {\bibfield  {journal} {\bibinfo  {journal} {Opt. Commun.}\
  }\textbf {\bibinfo {volume} {271}},\ \bibinfo {pages} {532} (\bibinfo {year}
  {2007})}\BibitemShut {NoStop}%
\bibitem [{\citenamefont {Kolokolov}\ and\ \citenamefont
  {Monovskaya}(2013)}]{Kolokolov2013}%
  \BibitemOpen
  \bibfield  {author} {\bibinfo {author} {\bibfnamefont {Y.}~\bibnamefont
  {Kolokolov}}\ and\ \bibinfo {author} {\bibfnamefont {A.}~\bibnamefont
  {Monovskaya}},\ }\href@noop {} {\bibfield  {journal} {\bibinfo  {journal}
  {Int. J. Bif. and Chaos}\ }\textbf {\bibinfo {volume} {23}},\ \bibinfo
  {pages} {1350063} (\bibinfo {year} {2013})}\BibitemShut {NoStop}%
\bibitem [{\citenamefont {Grebogi}\ \emph {et~al.}(1983)\citenamefont
  {Grebogi}, \citenamefont {Ott},\ and\ \citenamefont {Yorke}}]{Grebogi1983}%
  \BibitemOpen
  \bibfield  {author} {\bibinfo {author} {\bibfnamefont {C.}~\bibnamefont
  {Grebogi}}, \bibinfo {author} {\bibfnamefont {E.}~\bibnamefont {Ott}}, \ and\
  \bibinfo {author} {\bibfnamefont {J.~A.}\ \bibnamefont {Yorke}},\ }\href@noop
  {} {\bibfield  {journal} {\bibinfo  {journal} {Physica D}\ }\textbf {\bibinfo
  {volume} {7}},\ \bibinfo {pages} {181} (\bibinfo {year} {1983})}\BibitemShut
  {NoStop}%
\bibitem [{\citenamefont {Baptista}\ \emph {et~al.}(2003)\citenamefont
  {Baptista}, \citenamefont {Grebogi},\ and\ \citenamefont
  {Barreto}}]{Murilo2003}%
  \BibitemOpen
  \bibfield  {author} {\bibinfo {author} {\bibfnamefont {M.~S.}\ \bibnamefont
  {Baptista}}, \bibinfo {author} {\bibfnamefont {C.}~\bibnamefont {Grebogi}}, \
  and\ \bibinfo {author} {\bibfnamefont {E.}~\bibnamefont {Barreto}},\
  }\href@noop {} {\bibfield  {journal} {\bibinfo  {journal} {Int. J. of Bif.
  and Chaos}\ }\textbf {\bibinfo {volume} {13}},\ \bibinfo {pages} {2681}
  (\bibinfo {year} {2003})}\BibitemShut {NoStop}%
\bibitem [{\citenamefont {Fa\c{c}anha}\ \emph {et~al.}(2013)\citenamefont
  {Fa\c{c}anha}, \citenamefont {Oldeman},\ and\ \citenamefont
  {Glass}}]{Façanha2013}%
  \BibitemOpen
  \bibfield  {author} {\bibinfo {author} {\bibfnamefont {W.}~\bibnamefont
  {Fa\c{c}anha}}, \bibinfo {author} {\bibfnamefont {B.}~\bibnamefont
  {Oldeman}}, \ and\ \bibinfo {author} {\bibfnamefont {L.}~\bibnamefont
  {Glass}},\ }\href@noop {} {\bibfield  {journal} {\bibinfo  {journal} {Phys.
  Lett. A}\ }\textbf {\bibinfo {volume} {377}},\ \bibinfo {pages} {1264}
  (\bibinfo {year} {2013})}\BibitemShut {NoStop}%
\bibitem [{\citenamefont {Bonatto}\ \emph {et~al.}(2005)\citenamefont
  {Bonatto}, \citenamefont {Garreau},\ and\ \citenamefont
  {Gallas}}]{Bonatto2005}%
  \BibitemOpen
  \bibfield  {author} {\bibinfo {author} {\bibfnamefont {C.}~\bibnamefont
  {Bonatto}}, \bibinfo {author} {\bibfnamefont {J.~C.}\ \bibnamefont
  {Garreau}}, \ and\ \bibinfo {author} {\bibfnamefont {J.~A.~C.}\ \bibnamefont
  {Gallas}},\ }\href@noop {} {\bibfield  {journal} {\bibinfo  {journal} {Phys.
  Rev. Lett.}\ }\textbf {\bibinfo {volume} {95}},\ \bibinfo {pages} {143905}
  (\bibinfo {year} {2005})}\BibitemShut {NoStop}%
\bibitem [{\citenamefont {Albuquerque}\ and\ \citenamefont
  {Rech}(2012)}]{Albuquerque2012}%
  \BibitemOpen
  \bibfield  {author} {\bibinfo {author} {\bibfnamefont {H.~A.}\ \bibnamefont
  {Albuquerque}}\ and\ \bibinfo {author} {\bibfnamefont {P.~C.}\ \bibnamefont
  {Rech}},\ }\href@noop {} {\bibfield  {journal} {\bibinfo  {journal} {Int. J.
  Circ. Theor. Appl.}\ }\textbf {\bibinfo {volume} {40}},\ \bibinfo {pages}
  {189} (\bibinfo {year} {2012})}\BibitemShut {NoStop}%
\bibitem [{\citenamefont {Slipantschuk}\ \emph {et~al.}(2010)\citenamefont
  {Slipantschuk}, \citenamefont {Ullner}, \citenamefont {Baptista},
  \citenamefont {Zeineddine},\ and\ \citenamefont {Thiel}}]{Slipantschuk2010}%
  \BibitemOpen
  \bibfield  {author} {\bibinfo {author} {\bibfnamefont {J.}~\bibnamefont
  {Slipantschuk}}, \bibinfo {author} {\bibfnamefont {E.}~\bibnamefont
  {Ullner}}, \bibinfo {author} {\bibfnamefont {M.~S.}\ \bibnamefont
  {Baptista}}, \bibinfo {author} {\bibfnamefont {M.}~\bibnamefont
  {Zeineddine}}, \ and\ \bibinfo {author} {\bibfnamefont {M.}~\bibnamefont
  {Thiel}},\ }\href@noop {} {\bibfield  {journal} {\bibinfo  {journal} {Chaos}\
  }\textbf {\bibinfo {volume} {20}},\ \bibinfo {pages} {045117} (\bibinfo
  {year} {2010})}\BibitemShut {NoStop}%
\bibitem [{\citenamefont {Medeiros}\ \emph {et~al.}(2010)\citenamefont
  {Medeiros}, \citenamefont {de~Souza}, \citenamefont {Medrano-T},\ and\
  \citenamefont {Caldas}}]{Medeiros2010}%
  \BibitemOpen
  \bibfield  {author} {\bibinfo {author} {\bibfnamefont {E.~S.}\ \bibnamefont
  {Medeiros}}, \bibinfo {author} {\bibfnamefont {S.~L.~T.}\ \bibnamefont
  {de~Souza}}, \bibinfo {author} {\bibfnamefont {R.~O.}\ \bibnamefont
  {Medrano-T}}, \ and\ \bibinfo {author} {\bibfnamefont {I.~L.}\ \bibnamefont
  {Caldas}},\ }\href@noop {} {\bibfield  {journal} {\bibinfo  {journal} {Phys.
  Lett. A}\ }\textbf {\bibinfo {volume} {374}},\ \bibinfo {pages} {2628}
  (\bibinfo {year} {2010})}\BibitemShut {NoStop}%
\bibitem [{\citenamefont {Baptista}\ and\ \citenamefont
  {Caldas}(1996)}]{Murilo1996}%
  \BibitemOpen
  \bibfield  {author} {\bibinfo {author} {\bibfnamefont {M.~S.}\ \bibnamefont
  {Baptista}}\ and\ \bibinfo {author} {\bibfnamefont {I.~L.}\ \bibnamefont
  {Caldas}},\ }\href@noop {} {\bibfield  {journal} {\bibinfo  {journal} {Chaos,
  Solitons and Fractals}\ }\textbf {\bibinfo {volume} {7}},\ \bibinfo {pages}
  {325} (\bibinfo {year} {1996})}\BibitemShut {NoStop}%
\bibitem [{\citenamefont {Gallas}(2010)}]{Gallas2010}%
  \BibitemOpen
  \bibfield  {author} {\bibinfo {author} {\bibfnamefont {J.~A.~C.}\
  \bibnamefont {Gallas}},\ }\href@noop {} {\bibfield  {journal} {\bibinfo
  {journal} {Int. J. Bif. and Chaos}\ }\textbf {\bibinfo {volume} {20}},\
  \bibinfo {pages} {197} (\bibinfo {year} {2010})}\BibitemShut {NoStop}%
\bibitem [{\citenamefont {Bonatto}\ and\ \citenamefont
  {Gallas}(2008{\natexlab{a}})}]{Bonatto2008b}%
  \BibitemOpen
  \bibfield  {author} {\bibinfo {author} {\bibfnamefont {C.}~\bibnamefont
  {Bonatto}}\ and\ \bibinfo {author} {\bibfnamefont {J.~A.~C.}\ \bibnamefont
  {Gallas}},\ }\href@noop {} {\bibfield  {journal} {\bibinfo  {journal} {Phil.
  Trans. R. Soc. A}\ }\textbf {\bibinfo {volume} {366}},\ \bibinfo {pages}
  {505} (\bibinfo {year} {2008}{\natexlab{a}})}\BibitemShut {NoStop}%
\bibitem [{\citenamefont {Belair}\ and\ \citenamefont
  {Glass}(1983)}]{Glass1983}%
  \BibitemOpen
  \bibfield  {author} {\bibinfo {author} {\bibfnamefont {J.}~\bibnamefont
  {Belair}}\ and\ \bibinfo {author} {\bibfnamefont {L.}~\bibnamefont {Glass}},\
  }\href@noop {} {\bibfield  {journal} {\bibinfo  {journal} {Phys. Lett. A}\
  }\textbf {\bibinfo {volume} {96}},\ \bibinfo {pages} {113} (\bibinfo {year}
  {1983})}\BibitemShut {NoStop}%
\bibitem [{\citenamefont {Bonatto}\ and\ \citenamefont
  {Gallas}(2007)}]{Bonatto2007}%
  \BibitemOpen
  \bibfield  {author} {\bibinfo {author} {\bibfnamefont {C.}~\bibnamefont
  {Bonatto}}\ and\ \bibinfo {author} {\bibfnamefont {J.~A.~C.}\ \bibnamefont
  {Gallas}},\ }\href@noop {} {\bibfield  {journal} {\bibinfo  {journal} {Phys.
  Rev. E}\ }\textbf {\bibinfo {volume} {75}},\ \bibinfo {pages} {055204(R)}
  (\bibinfo {year} {2007})}\BibitemShut {NoStop}%
\bibitem [{\citenamefont {Medeiros}\ \emph {et~al.}(2013)\citenamefont
  {Medeiros}, \citenamefont {Medrano-T}, \citenamefont {Caldas},\ and\
  \citenamefont {de~Souza}}]{Medeiros2013}%
  \BibitemOpen
  \bibfield  {author} {\bibinfo {author} {\bibfnamefont {E.~S.}\ \bibnamefont
  {Medeiros}}, \bibinfo {author} {\bibfnamefont {R.~O.}\ \bibnamefont
  {Medrano-T}}, \bibinfo {author} {\bibfnamefont {I.~L.}\ \bibnamefont
  {Caldas}}, \ and\ \bibinfo {author} {\bibfnamefont {S.~L.~T.}\ \bibnamefont
  {de~Souza}},\ }\href@noop {} {\bibfield  {journal} {\bibinfo  {journal}
  {Phys. Lett. A}\ }\textbf {\bibinfo {volume} {377}},\ \bibinfo {pages} {628}
  (\bibinfo {year} {2013})}\BibitemShut {NoStop}%
\bibitem [{\citenamefont {Gaspard}\ and\ \citenamefont
  {Nicolis}(1983)}]{Gaspard1982}%
  \BibitemOpen
  \bibfield  {author} {\bibinfo {author} {\bibfnamefont {P.}~\bibnamefont
  {Gaspard}}\ and\ \bibinfo {author} {\bibfnamefont {G.}~\bibnamefont
  {Nicolis}},\ }\href@noop {} {\bibfield  {journal} {\bibinfo  {journal} {J.
  Stat. Phys.}\ }\textbf {\bibinfo {volume} {31}},\ \bibinfo {pages} {499}
  (\bibinfo {year} {1983})}\BibitemShut {NoStop}%
\bibitem [{\citenamefont {Gaspard}\ \emph {et~al.}(1984)\citenamefont
  {Gaspard}, \citenamefont {Kapral},\ and\ \citenamefont
  {Nicolis}}]{Gaspard1984}%
  \BibitemOpen
  \bibfield  {author} {\bibinfo {author} {\bibfnamefont {P.}~\bibnamefont
  {Gaspard}}, \bibinfo {author} {\bibfnamefont {R.}~\bibnamefont {Kapral}}, \
  and\ \bibinfo {author} {\bibfnamefont {G.}~\bibnamefont {Nicolis}},\
  }\href@noop {} {\bibfield  {journal} {\bibinfo  {journal} {J. Stat. Phys.}\
  }\textbf {\bibinfo {volume} {35}},\ \bibinfo {pages} {697} (\bibinfo {year}
  {1984})}\BibitemShut {NoStop}%
\bibitem [{\citenamefont {Medrano-T}\ and\ \citenamefont
  {Caldas}(2010)}]{Rene2010}%
  \BibitemOpen
  \bibfield  {author} {\bibinfo {author} {\bibfnamefont {R.~O.}\ \bibnamefont
  {Medrano-T}}\ and\ \bibinfo {author} {\bibfnamefont {I.~L.}\ \bibnamefont
  {Caldas}},\ }\href@noop {} {\bibfield  {journal} {\bibinfo  {journal} {ArXiv
  e-prints}\ } (\bibinfo {year} {2010})},\ \Eprint
  {http://arxiv.org/abs/1012.2241} {1012.2241} \BibitemShut {NoStop}%
\bibitem [{\citenamefont {Vitolo}\ \emph {et~al.}(2011)\citenamefont {Vitolo},
  \citenamefont {Glendinning},\ and\ \citenamefont {Gallas}}]{Vitolo2011}%
  \BibitemOpen
  \bibfield  {author} {\bibinfo {author} {\bibfnamefont {R.}~\bibnamefont
  {Vitolo}}, \bibinfo {author} {\bibfnamefont {P.}~\bibnamefont {Glendinning}},
  \ and\ \bibinfo {author} {\bibfnamefont {J.~A.~C.}\ \bibnamefont {Gallas}},\
  }\href@noop {} {\bibfield  {journal} {\bibinfo  {journal} {Phys. Rev. E}\
  }\textbf {\bibinfo {volume} {84}},\ \bibinfo {pages} {016216} (\bibinfo
  {year} {2011})}\BibitemShut {NoStop}%
\bibitem [{\citenamefont {Barrio}\ \emph {et~al.}(2011)\citenamefont {Barrio},
  \citenamefont {Blesa}, \citenamefont {Serrano},\ and\ \citenamefont
  {Shilnikov}}]{Barrio2011}%
  \BibitemOpen
  \bibfield  {author} {\bibinfo {author} {\bibfnamefont {R.}~\bibnamefont
  {Barrio}}, \bibinfo {author} {\bibfnamefont {F.}~\bibnamefont {Blesa}},
  \bibinfo {author} {\bibfnamefont {S.}~\bibnamefont {Serrano}}, \ and\
  \bibinfo {author} {\bibfnamefont {A.}~\bibnamefont {Shilnikov}},\ }\href@noop
  {} {\bibfield  {journal} {\bibinfo  {journal} {Phys. Rev. E}\ }\textbf
  {\bibinfo {volume} {84}},\ \bibinfo {pages} {035201(R)} (\bibinfo {year}
  {2011})}\BibitemShut {NoStop}%
\bibitem [{\citenamefont {Barrio}\ \emph {et~al.}(2012)\citenamefont {Barrio},
  \citenamefont {Blesa},\ and\ \citenamefont {Serrano}}]{Barrio2012}%
  \BibitemOpen
  \bibfield  {author} {\bibinfo {author} {\bibfnamefont {R.}~\bibnamefont
  {Barrio}}, \bibinfo {author} {\bibfnamefont {F.}~\bibnamefont {Blesa}}, \
  and\ \bibinfo {author} {\bibfnamefont {S.}~\bibnamefont {Serrano}},\
  }\href@noop {} {\bibfield  {journal} {\bibinfo  {journal} {Phys. Rev. Lett.}\
  }\textbf {\bibinfo {volume} {108}},\ \bibinfo {pages} {214102} (\bibinfo
  {year} {2012})}\BibitemShut {NoStop}%
\bibitem [{\citenamefont {Bonatto}\ and\ \citenamefont
  {Gallas}(2008{\natexlab{b}})}]{Bonatto2008}%
  \BibitemOpen
  \bibfield  {author} {\bibinfo {author} {\bibfnamefont {C.}~\bibnamefont
  {Bonatto}}\ and\ \bibinfo {author} {\bibfnamefont {J.~A.~C.}\ \bibnamefont
  {Gallas}},\ }\href@noop {} {\bibfield  {journal} {\bibinfo  {journal} {Phys.
  Rev. Lett.}\ }\textbf {\bibinfo {volume} {101}},\ \bibinfo {pages} {0541011}
  (\bibinfo {year} {2008}{\natexlab{b}})}\BibitemShut {NoStop}%
\bibitem [{\citenamefont {Albuquerque}\ \emph {et~al.}(2008)\citenamefont
  {Albuquerque}, \citenamefont {Rubinger},\ and\ \citenamefont
  {Rech}}]{Holokx2008}%
  \BibitemOpen
  \bibfield  {author} {\bibinfo {author} {\bibfnamefont {H.~A.}\ \bibnamefont
  {Albuquerque}}, \bibinfo {author} {\bibfnamefont {R.~M.}\ \bibnamefont
  {Rubinger}}, \ and\ \bibinfo {author} {\bibfnamefont {P.~C.}\ \bibnamefont
  {Rech}},\ }\href@noop {} {\bibfield  {journal} {\bibinfo  {journal} {Phys.
  Lett. A}\ }\textbf {\bibinfo {volume} {372}},\ \bibinfo {pages} {4793}
  (\bibinfo {year} {2008})}\BibitemShut {NoStop}%
\bibitem [{\citenamefont {Maranh{\~a}o}\ \emph {et~al.}(2008)\citenamefont
  {Maranh{\~a}o}, \citenamefont {Baptista}, \citenamefont {Sartoreli},\ and\
  \citenamefont {Caldas}}]{Maranhao2008}%
  \BibitemOpen
  \bibfield  {author} {\bibinfo {author} {\bibfnamefont {D.~M.}\ \bibnamefont
  {Maranh{\~a}o}}, \bibinfo {author} {\bibfnamefont {M.~S.}\ \bibnamefont
  {Baptista}}, \bibinfo {author} {\bibfnamefont {J.~C.}\ \bibnamefont
  {Sartoreli}}, \ and\ \bibinfo {author} {\bibfnamefont {I.~L.}\ \bibnamefont
  {Caldas}},\ }\href@noop {} {\bibfield  {journal} {\bibinfo  {journal} {Phys.
  Rev. E}\ }\textbf {\bibinfo {volume} {77}},\ \bibinfo {pages} {037202}
  (\bibinfo {year} {2008})}\BibitemShut {NoStop}%
\bibitem [{\citenamefont {Stoop}\ \emph {et~al.}(2010)\citenamefont {Stoop},
  \citenamefont {Benner},\ and\ \citenamefont {Uwate}}]{Stoop2010}%
  \BibitemOpen
  \bibfield  {author} {\bibinfo {author} {\bibfnamefont {R.}~\bibnamefont
  {Stoop}}, \bibinfo {author} {\bibfnamefont {P.}~\bibnamefont {Benner}}, \
  and\ \bibinfo {author} {\bibfnamefont {Y.}~\bibnamefont {Uwate}},\
  }\href@noop {} {\bibfield  {journal} {\bibinfo  {journal} {Phys. Rev. Lett.}\
  }\textbf {\bibinfo {volume} {105}},\ \bibinfo {pages} {074102} (\bibinfo
  {year} {2010})}\BibitemShut {NoStop}%
\bibitem [{\citenamefont {Cabeza}\ \emph {et~al.}(2013)\citenamefont {Cabeza},
  \citenamefont {Briozzo}, \citenamefont {Garcia}, \citenamefont {Freire},
  \citenamefont {Marti},\ and\ \citenamefont {Gallas}}]{Cabeza2013}%
  \BibitemOpen
  \bibfield  {author} {\bibinfo {author} {\bibfnamefont {C.}~\bibnamefont
  {Cabeza}}, \bibinfo {author} {\bibfnamefont {C.~A.}\ \bibnamefont {Briozzo}},
  \bibinfo {author} {\bibfnamefont {R.}~\bibnamefont {Garcia}}, \bibinfo
  {author} {\bibfnamefont {J.~G.}\ \bibnamefont {Freire}}, \bibinfo {author}
  {\bibfnamefont {A.~C.}\ \bibnamefont {Marti}}, \ and\ \bibinfo {author}
  {\bibfnamefont {J.~A.}\ \bibnamefont {Gallas}},\ }\href@noop {} {\bibfield
  {journal} {\bibinfo  {journal} {Chaos, Solitons and Fractals}\ }\textbf
  {\bibinfo {volume} {52}},\ \bibinfo {pages} {59} (\bibinfo {year}
  {2013})}\BibitemShut {NoStop}%
\bibitem [{\citenamefont {Hunt}\ and\ \citenamefont {Ott}(1997)}]{Hunt1997}%
  \BibitemOpen
  \bibfield  {author} {\bibinfo {author} {\bibfnamefont {B.}~\bibnamefont
  {Hunt}}\ and\ \bibinfo {author} {\bibfnamefont {E.}~\bibnamefont {Ott}},\
  }\href@noop {} {\bibfield  {journal} {\bibinfo  {journal} {J. Phys. A: Math.
  Gen.}\ }\textbf {\bibinfo {volume} {30}},\ \bibinfo {pages} {7067} (\bibinfo
  {year} {1997})}\BibitemShut {NoStop}%
\bibitem [{\citenamefont {Eykholt}\ and\ \citenamefont
  {Umberger}(1986)}]{Umberger1986}%
  \BibitemOpen
  \bibfield  {author} {\bibinfo {author} {\bibfnamefont {R.}~\bibnamefont
  {Eykholt}}\ and\ \bibinfo {author} {\bibfnamefont {D.~K.}\ \bibnamefont
  {Umberger}},\ }\href@noop {} {\bibfield  {journal} {\bibinfo  {journal}
  {Phys. Rev. Lett.}\ }\textbf {\bibinfo {volume} {57}},\ \bibinfo {pages}
  {2333} (\bibinfo {year} {1986})}\BibitemShut {NoStop}%
\bibitem [{\citenamefont {Farmer}(1985)}]{Farmer1985}%
  \BibitemOpen
  \bibfield  {author} {\bibinfo {author} {\bibfnamefont {J.~D.}\ \bibnamefont
  {Farmer}},\ }\href@noop {} {\bibfield  {journal} {\bibinfo  {journal} {Phys.
  Rev. lett.}\ }\textbf {\bibinfo {volume} {55}},\ \bibinfo {pages} {351}
  (\bibinfo {year} {1985})}\BibitemShut {NoStop}%
\bibitem [{\citenamefont {Jakobson}(1981)}]{Jakobson1981}%
  \BibitemOpen
  \bibfield  {author} {\bibinfo {author} {\bibfnamefont {M.~V.}\ \bibnamefont
  {Jakobson}},\ }\href@noop {} {\bibfield  {journal} {\bibinfo  {journal}
  {Commun. Math. Phys.}\ }\textbf {\bibinfo {volume} {81}},\ \bibinfo {pages}
  {39} (\bibinfo {year} {1981})}\BibitemShut {NoStop}%
\bibitem [{\citenamefont {Gallas}(1993)}]{Gallas1993}%
  \BibitemOpen
  \bibfield  {author} {\bibinfo {author} {\bibfnamefont {J.~A.~C.}\
  \bibnamefont {Gallas}},\ }\href@noop {} {\bibfield  {journal} {\bibinfo
  {journal} {Phys. Rev. Lett.}\ }\textbf {\bibinfo {volume} {70}},\ \bibinfo
  {pages} {2714} (\bibinfo {year} {1993})}\BibitemShut {NoStop}%
\bibitem [{\citenamefont {Lorenz}(2008)}]{Lorenz2008}%
  \BibitemOpen
  \bibfield  {author} {\bibinfo {author} {\bibfnamefont {E.~N.}\ \bibnamefont
  {Lorenz}},\ }\href@noop {} {\bibfield  {journal} {\bibinfo  {journal}
  {Physica D}\ }\textbf {\bibinfo {volume} {13}},\ \bibinfo {pages} {1689}
  (\bibinfo {year} {2008})}\BibitemShut {NoStop}%
\bibitem [{\citenamefont {Viana}\ \emph {et~al.}(2010)\citenamefont {Viana},
  \citenamefont {Rubinger}, \citenamefont {Albuquerque}, \citenamefont
  {de~Oliveira},\ and\ \citenamefont {Ribeiro}}]{Viana2010}%
  \BibitemOpen
  \bibfield  {author} {\bibinfo {author} {\bibfnamefont {E.~R.}\ \bibnamefont
  {Viana}}, \bibinfo {author} {\bibfnamefont {R.~M.}\ \bibnamefont {Rubinger}},
  \bibinfo {author} {\bibfnamefont {H.~A.}\ \bibnamefont {Albuquerque}},
  \bibinfo {author} {\bibfnamefont {A.~G.}\ \bibnamefont {de~Oliveira}}, \ and\
  \bibinfo {author} {\bibfnamefont {G.~M.}\ \bibnamefont {Ribeiro}},\
  }\href@noop {} {\bibfield  {journal} {\bibinfo  {journal} {Chaos}\ }\textbf
  {\bibinfo {volume} {20}},\ \bibinfo {pages} {0231101} (\bibinfo {year}
  {2010})}\BibitemShut {NoStop}%
\bibitem [{Not()}]{Nota1}%
  \BibitemOpen
  \href@noop {} {}\bibinfo {note} {The window width has been proposed to be
  dependent of the orbital period $p$ and of the topological entropy $h_T$ of
  the nearby chaotic attractors as an exponential law for quadratic maps
  \cite{Hunt1997}, we remark that this previous result is for periodic windows
  independent of any choice of accumulation sequences. In addition, notice that
  $h_T$ change along the sequence. The functional form of $h_T$ as function of
  the parameters along the accumulation sequence is unknown. Therefore, our
  measurements do not exclude its validity for the dynamical systems considered
  here.}\BibitemShut {Stop}%
\bibitem [{\citenamefont {Grebogi}\ \emph {et~al.}(1985)\citenamefont
  {Grebogi}, \citenamefont {McDonald}, \citenamefont {Ott},\ and\ \citenamefont
  {Yorke}}]{Grebogi1985}%
  \BibitemOpen
  \bibfield  {author} {\bibinfo {author} {\bibfnamefont {C.}~\bibnamefont
  {Grebogi}}, \bibinfo {author} {\bibfnamefont {S.~W.}\ \bibnamefont
  {McDonald}}, \bibinfo {author} {\bibfnamefont {E.}~\bibnamefont {Ott}}, \
  and\ \bibinfo {author} {\bibfnamefont {J.~A.}\ \bibnamefont {Yorke}},\
  }\href@noop {} {\bibfield  {journal} {\bibinfo  {journal} {Phys. Lett. A}\
  }\textbf {\bibinfo {volume} {110}},\ \bibinfo {pages} {01} (\bibinfo {year}
  {1985})}\BibitemShut {NoStop}%
\bibitem [{\citenamefont {McDonald}\ \emph {et~al.}(1984)\citenamefont
  {McDonald}, \citenamefont {Grebogi}, \citenamefont {Ott},\ and\ \citenamefont
  {Yorke}}]{McDonald1984}%
  \BibitemOpen
  \bibfield  {author} {\bibinfo {author} {\bibfnamefont {S.~W.}\ \bibnamefont
  {McDonald}}, \bibinfo {author} {\bibfnamefont {C.}~\bibnamefont {Grebogi}},
  \bibinfo {author} {\bibfnamefont {E.}~\bibnamefont {Ott}}, \ and\ \bibinfo
  {author} {\bibfnamefont {J.~A.}\ \bibnamefont {Yorke}},\ }\href@noop {}
  {\bibfield  {journal} {\bibinfo  {journal} {Physica D}\ }\textbf {\bibinfo
  {volume} {17}},\ \bibinfo {pages} {125} (\bibinfo {year} {1984})}\BibitemShut
  {NoStop}%
\bibitem [{\citenamefont {Scheffczyk}\ \emph {et~al.}(1991)\citenamefont
  {Scheffczyk}, \citenamefont {Parlitz}, \citenamefont {Kurz}, \citenamefont
  {Konp},\ and\ \citenamefont {Lauterborn}}]{Scheffczyk1991}%
  \BibitemOpen
  \bibfield  {author} {\bibinfo {author} {\bibfnamefont {C.}~\bibnamefont
  {Scheffczyk}}, \bibinfo {author} {\bibfnamefont {U.}~\bibnamefont {Parlitz}},
  \bibinfo {author} {\bibfnamefont {T.}~\bibnamefont {Kurz}}, \bibinfo {author}
  {\bibfnamefont {W.}~\bibnamefont {Konp}}, \ and\ \bibinfo {author}
  {\bibfnamefont {W.}~\bibnamefont {Lauterborn}},\ }\href@noop {} {\bibfield
  {journal} {\bibinfo  {journal} {Phys. Rev. A}\ }\textbf {\bibinfo {volume}
  {43}},\ \bibinfo {pages} {6495} (\bibinfo {year} {1991})}\BibitemShut
  {NoStop}%
\bibitem [{\citenamefont {Jensen}\ \emph {et~al.}(1983)\citenamefont {Jensen},
  \citenamefont {Bak},\ and\ \citenamefont {Bohr}}]{Perbak1983}%
  \BibitemOpen
  \bibfield  {author} {\bibinfo {author} {\bibfnamefont {M.~H.}\ \bibnamefont
  {Jensen}}, \bibinfo {author} {\bibfnamefont {P.}~\bibnamefont {Bak}}, \ and\
  \bibinfo {author} {\bibfnamefont {T.}~\bibnamefont {Bohr}},\ }\href@noop {}
  {\bibfield  {journal} {\bibinfo  {journal} {Phys. Rev. Lett.}\ }\textbf
  {\bibinfo {volume} {50}},\ \bibinfo {pages} {1637} (\bibinfo {year}
  {1983})}\BibitemShut {NoStop}%
\bibitem [{\citenamefont {Yorke}\ \emph {et~al.}(1985)\citenamefont {Yorke},
  \citenamefont {Grebogi}, \citenamefont {Ott},\ and\ \citenamefont
  {Tedeschini-Lalli}}]{Yorke1985}%
  \BibitemOpen
  \bibfield  {author} {\bibinfo {author} {\bibfnamefont {J.~A.}\ \bibnamefont
  {Yorke}}, \bibinfo {author} {\bibfnamefont {C.}~\bibnamefont {Grebogi}},
  \bibinfo {author} {\bibfnamefont {E.}~\bibnamefont {Ott}}, \ and\ \bibinfo
  {author} {\bibfnamefont {L.}~\bibnamefont {Tedeschini-Lalli}},\ }\href@noop
  {} {\bibfield  {journal} {\bibinfo  {journal} {Phys. Rev. Lett.}\ }\textbf
  {\bibinfo {volume} {54}},\ \bibinfo {pages} {1095} (\bibinfo {year}
  {1985})}\BibitemShut {NoStop}%
\bibitem [{\citenamefont {Glass}\ and\ \citenamefont
  {Perez}(1982)}]{Glass1982}%
  \BibitemOpen
  \bibfield  {author} {\bibinfo {author} {\bibfnamefont {L.}~\bibnamefont
  {Glass}}\ and\ \bibinfo {author} {\bibfnamefont {R.}~\bibnamefont {Perez}},\
  }\href@noop {} {\bibfield  {journal} {\bibinfo  {journal} {Phys. Rev. Lett.}\
  }\textbf {\bibinfo {volume} {48}},\ \bibinfo {pages} {1772} (\bibinfo {year}
  {1982})}\BibitemShut {NoStop}%
\bibitem [{\citenamefont {Ecke}\ \emph {et~al.}(1989)\citenamefont {Ecke},
  \citenamefont {Farmer},\ and\ \citenamefont {Umberger}}]{Ecke1989}%
  \BibitemOpen
  \bibfield  {author} {\bibinfo {author} {\bibfnamefont {R.~E.}\ \bibnamefont
  {Ecke}}, \bibinfo {author} {\bibfnamefont {J.~D.}\ \bibnamefont {Farmer}}, \
  and\ \bibinfo {author} {\bibfnamefont {D.~K.}\ \bibnamefont {Umberger}},\
  }\href@noop {} {\bibfield  {journal} {\bibinfo  {journal} {Nonlinearity}\
  }\textbf {\bibinfo {volume} {2}},\ \bibinfo {pages} {175} (\bibinfo {year}
  {1989})}\BibitemShut {NoStop}%
\end{thebibliography}
\end{document}